\documentstyle[12pt,epsf,epsfig,a4]{article}
\begin{document}
\begin {titlepage}
\begin{flushleft}
FSUJ TPI QO-7/98
\end{flushleft}
\begin{flushright}
May, 1998
\end{flushright}
\vspace{20mm}
\begin{center}
{\Large {\bf Generating and monitoring Schr\"odinger cats in
conditional measurement on a beam splitter}
\\[3ex]
\Large M. Dakna, J. Clausen, L. Kn\"oll, D.--G.  Welsch}\\[3ex]
Friedrich-Schiller-Universit\"{a}t Jena,
Theoretisch-Physikalisches Institut
\\
Max-Wien-Platz 1, D-07743 Jena, Germany\\[2.ex] 
\vspace{25mm}
\end{center}
\begin{center}
\bf{Abstract}
\end{center}
Preparation of Schr\"odinger-cat-like states via conditional 
output measurement on a beam splitter is studied. In the scheme,
a mode prepared in a squeezed vacuum is mixed with a mode prepared in a 
Fock state and photocounting is performed in one of the 
output channels of the beam splitter. 
In this way the mode in the other output channel is prepared in 
a Schr\"odinger-cat-like state that is either 
a photon-subtracted or a photon-added Jacobi polynomial squeezed vacuum 
state, depending upon the difference between the number of photons in the 
input Fock state and the number of photons in the output Fock state 
onto which it is projected. Two possible photocounting schemes are
considered, and the problem of monitoring cats that are ``hidden''
in a statistical mixture of states is studied.     
\end{titlepage}
\section{Introduction}
\label{sec1}
The interference of probability amplitudes is one of the most specific 
features of quantum theory and it has been discussed since the early days 
of quantum mechanics. The famous Schr\"odinger-cat-like states are a typical 
example. The cat stands for a macroscopic object which may be in a 
superposition of states corresponding to macroscopically distinguishable 
beings (living and dead) \cite{Schroedinger1}. However the experimental 
demonstration of quantum interference effects for macroscopic systems is 
very difficult due to the irreversible interaction of such a system with 
its environment \cite{Zurek1}. Recent experimental progress has
rendered it possible to generate Schr\"odinger-cat-like states on a
mesoscopic scale, as it was successfully demonstrated in neutron interferometry 
\cite{Rauch} and atom optics in a trap \cite {Monroe1}. 

Despite the large body of work, Schr\"odinger-cat-like states
of optical fields have not been observed so far. Recently
a sophisticated method for demonstrating Schr\"odinger-cat-like 
states of traveling optical fields by inferring them from noisy 
data has been proposed \cite{DARIANO1}. The method is based on the
experimental scheme proposed in \cite{Song1} (see also \cite{Yurke1}), 
and it uses appropriate data processing in order to calculate 
the intrinsic (undistorted) state from the (distorted) state 
actually produced and detected.

Recently we have shown that Schr\"odinger-cat-like states 
can also be produced by conditional measurement on a beam splitter.
In particular, when a mode prepared in a squeezed vacuum is mixed with 
an ordinary vacuum and a (nonzero) photon-number measurement is performed 
in one of the output channels of the beam splitter, then the mode in the 
other output channel is prepared in a Schr\"odinger-cat-like state 
\cite{Dakna1}. The scheme can be extended to 
the more general case when the mode prepared in the squeezed vacuum
is mixed with a mode prepared in an arbitrary $n$-photon Fock state
and the photon-number measurement yields an arbitrary number $m$. 
The conditional output states 
are either photon-subtracted ($n$ $\!<$ $\!m$) or photon-added 
($n$ $\!>$ $\!m$) Jacobi polynomial squeezed vacuum states, i.e., 
states that are obtained by ($|n$ $\!-$ $\!m|$ times) repeated application 
of either the photon destruction operator or the photon creation operator, 
respectively, to a Jacobi polynomial squeezed vacuum state \cite{Dakna2}. 
It is worth noting that the two classes of 
states -- similarly to the ordinary photon-subtracted 
\mbox{($n$ $\!=$ $\!0$)} \cite{Dakna1}
and photon-added ($m$ $\!=$ $\!0$) \cite{Dakna3} squeezed vacuum states 
-- represent Schr\"odinger-cat-like states.

In this paper we study the properties of these classes of
Schr\"odinger-cat-like states, with special emphasis on the experimental 
conditions. In fact, due to the extreme fragility of the quantum 
interferences the attempt to demonstrate them   
may fail if no rigorous compensation for the
detection losses is performed, at least in the conditional measurement. We
compare photon chopping \cite{Paul1} (considered in \cite{Dakna1}) with 
single-detector photocounting (considered in \cite{DARIANO1}) and give 
an analysis of the corresponding data processing algorithms.
Further, the effect of losses in homodyne detection of the
states produced is addressed. 

This paper is organized as follows. In Sec.~\ref{sec2} we outline 
the basic scheme for generating photon-subtracted Jacobi polynomial (PSJP)
and photon-added Jacobi polynomial (PAJP) squeezed vacuum states and 
briefly address their properties. In Sec.~\ref{sec3} we compare 
the two photocounting schemes for conditional measurement
and analyse the corresponding algorithms used for processing the 
experimental data. Finally, in Sec.~\ref{sec4} we give
a summary and some concluding remarks. 

\section{Basis scheme for generating the Schr\"odinger-cat-like states} 
\label{sec2}

The quantum description of  the input--output relations of a lossless beam
splitter are well known to obey  the $\rm SU(2)$ Lie algebra
\cite{Campos1}. In the Schr\"{o}dinger picture, 
the output-state density operator $\hat{\varrho}_{\rm out}$ can be
related to the input-state density operator $\hat{\varrho}_{\rm in}$ as
$\hat{\varrho}_{\rm out}\!=\!\hat{V}^{\dagger}\hat \varrho_{\rm in}\hat{V}$,
where $\hat{V}$ can be given by \cite{Campos1}
\begin{equation}
\hat{V} =
e^{-i(\varphi_{T}-\varphi_{R}) \hat{L}_{3}}
\, e^{-2i\theta \hat{L}_{2}}
\, e^{-i(\varphi_{T}+\varphi_{R}) \hat{L}_{3}},
\label{2.0}
\end{equation}
with
\begin{equation}
\hat{L}_{2} = \textstyle\frac{1}{2i}(\hat{a}_{1}^\dagger\hat{a}_{2}
-\hat{a}_{2}^\dagger\hat{a}_{1}), \quad
\hat{L}_{3} = \textstyle\frac{1}{2}(\hat{a}_{1}^\dagger\hat{a}_{1}
-\hat{a}_{2}^\dagger\hat{a}_{2}).
\label{2.1}
\end{equation}
Let us consider the experimental setup as depicted in Fig.~1. A 
radiation-field mode prepared in a quantum state $\hat{\varrho}_{{\rm in}1}$ 
is mixed at a beam splitter with another mode prepared in a 
Fock state $|n\rangle$, so that the input-state density operator reads as
\begin{equation}
\hat \varrho_{\rm in}(n) = \hat \varrho_{{\rm in}1}
\otimes |n\rangle_{2} \, _{2}\langle n|.
\label{2.3}
\end{equation}
The output-state density operator
$\hat{\varrho}_{\rm out}$ $\!\equiv$ $\!\hat{\varrho}_{\rm out}(n)$ 
can then be given by \cite{Dakna2}
\begin{eqnarray}
\lefteqn{
\hat \varrho _{\rm out}(n) = \frac{1}{|T|^{2n}}
\sum_{l=0}^{\infty}\sum_{m=0}^{\infty}
\sum_{k=0}^{n}\sum_{j=0}^{n} 
\frac{(-1)^{l+m}(R^*)^{m+j} R^{l+k}  }{ \sqrt{k!j!m!l!} }
\left[{n\!-\!k\!+\!m\choose m}\!{n\!-\!j\!+\!l\choose l} \right]^{\frac{1}{2}}
}
\nonumber \\ && \hspace{3ex}\times
\left[\! {n\choose k}\!{n\choose j}\right]^{\frac{1}{2}} 
T^{\hat{n}_{1}}
{\hat{a}_{1}}^{m}({\hat{a}_{1}^\dagger})^{k}
\hat{\varrho}_{{\rm in}1}\hat{a}_{1}^{j}
({\hat{a}_{1}^\dagger})^{l}(T^*)^{\hat{n}_{1}}
\!\otimes \!
|n\!-\!k\!+\!m\rangle_{2} \, _{2}\langle n\!-\!j\!+\!l|. 
\label{2.4}
\end{eqnarray}
   From Eq.~(\ref{2.4}) we see that the output modes are prepared
in a highly entangled quantum state in general. When the photon number 
of the mode in the second output channel is measured and $m$ photons are 
detected, then the mode in the first output channel is prepared in a 
quantum state whose density operator $\hat{\varrho}_{{\rm out}1}(n,m)$ reads 
\begin{eqnarray}
\hat{\varrho}_{{\rm out}1}(n,m)
= \frac{_{2}\langle m |\hat{\varrho}_{\rm out}(n) | m \rangle_{2}}
{{\rm Tr}_{1}\{ _{2}\langle m |
\hat{\varrho}_{\rm out}(n) | m \rangle_{2}\}} \,.
\label{2.5}
\end{eqnarray}
The probability of such an event is given by
\begin{eqnarray}
\lefteqn{
\hspace*{-2ex}
P(n,m)=
{\rm Tr}_{1}\{ _{2}\langle m |
\hat{\varrho}_{\rm out}{(n)} | m \rangle_{2} \}
}
\nonumber \\ && \hspace{-6ex}
=\!\frac{|R|^{-2\nu}n!}{|T|^{2m}m!}  
\sum_{j=\mu}^{n}\sum_{k=\mu}^{n}
(-|R|^2)^{j\!+\!k}\!{m\choose j\!-\!\nu}\!{m\choose k\!-\!\nu}\!
\sum_{p=\delta}^{\infty}\frac{p!|T|^{2p}}{(p\!+\!\nu)!}
\!{p\!+\!j\choose j}\!{p\!+\!k\choose k}
\langle p |\!\hat{\varrho}_{{\rm in}1}\! | p \rangle,
\label{2.6}
\end{eqnarray}
where the abbreviations
\begin{equation}
\nu = n - m, \quad
\mu=\max(0,\nu), \quad \delta=\mu-\nu
\end{equation}
have been used. 

Let us now consider the case when the first input mode is prepared in a
squeezed vacuum state $\hat{\varrho}_{{\rm in}1}\!
=\!\hat{S}(\xi)| 0 \rangle\langle 0  | \hat{S}^{\dagger}(\xi)$, where
\begin{eqnarray}
\hat{S}(\xi)| 0 \rangle =
\exp\!\left\{-\textstyle\frac{1}{2}
\left[\xi(\hat{a}_1^\dagger){^2} - \xi^{\ast}\hat{a}_1^{2}\right]
\right\} | 0 \rangle
=\,(1-|\kappa|^{2})^{\frac{1}{4}}\sum_{p=0}^{\infty}
\frac{[(2p)!]^{1/2}}{2^{p}\,p!}\,
\kappa^{p}|2p\rangle,
\label{2.7}
\end{eqnarray}
$\xi$ $\!=$ $\!|\xi|e^{i\varphi_{\xi}}$, $\kappa$ $\!=$ 
$\!-e^{i\varphi_{\xi}}\tanh|\xi|$.
Combining Eqs.~(\ref{2.4}) and (\ref{2.5}) and using Eq.~(\ref{2.7}), we 
find that the density operator of the output state reads \cite{Dakna3}
\begin{eqnarray}
\hat{\varrho}_{{\rm out}1}(n,m)\!=\!| \Psi_{n,m} 
\rangle \big\langle \Psi_{n,m}| ,
\label{2.8}
\end{eqnarray}
where 
\begin{eqnarray}
|\Psi_{n,m}\rangle\sim  
\left\{
\begin{array}{ll}
\hat{a}^{|\nu|}\, 
{\rm P}_n^{(|\nu|,\hat{n}-m)}(2|T|^2\!-\!1)
\hat{S}(\xi')|0\rangle 
& {\rm for} \, n\!<\!m, \\[1ex]                 
(\hat{a}^\dagger)^{\nu}
{\rm P}_m^{(\nu,\hat{n}-m)}(2|T|^2\!-\!1)
\hat{S}(\xi')|0\rangle
& {\rm for} \,  n\!>\!m                        
\end{array}\right.
\quad
\label{jacobi}
\end{eqnarray}
is a PSJP ($n$ $\!<$ $\!m$) or a PAJP ($n$ $\!>$ $\!m$) squeezed vacuum state
[${\rm P}_l^{(\alpha,\beta)}(z)$, Jacobi polynomial;
$\xi'$ $\!=$ $\!|\xi'|e^{i(\varphi_{\xi}+2\varphi_T)}$;
$\tanh|\xi'|$ $\!=$ $\!|T|^2\tanh|\xi|$].  
In the photon-number basis $|\Psi_{n,m}\rangle$ reads 
\begin{eqnarray}
\lefteqn{
\hspace*{2ex}
|\Psi_{n,m}\rangle
={\cal N}_{n,m}^{-1/2}
\sum_{k=\mu}^{n}\frac{(-|R|^2)^k}{(k-\nu)!}{n\choose k}
}
\nonumber \\ && \hspace{7ex}\times \,
\!\sum_{p=\mu}^{\infty}
\frac{(p!)^{-\frac{1}{2}}(p-\nu+k)!}
{\Gamma\!\left(\frac{p-\nu}{2}+1\right)}
{\textstyle\frac{1}{2}}\!\left\{1\!+\!(-1)^{p-\nu}\right\} 
\left({\textstyle\frac{1}{2}}\kappa'\right)^{(p-\nu)/2}  |p\rangle,
\label{2.9}
\end{eqnarray}
with $\kappa'$ $\!=$ $\!T^2\kappa$. From Eq.~(\ref{2.9}) we easily 
see that when the difference between the number $n$ of photons in the 
second input channel of the beam splitter and the number $m$ of photons 
detected in the second output channel, i.e., the parameter
$\nu$ $\!=$ $\!n$ $\!-$ $\!m$, is even (odd), then the mode in the first 
output channel is prepared in a PSJP or PAJP squeezed vacuum state 
$|\Psi_{n,m}\rangle$ that contains only Fock states with even (odd) numbers
of photons. Similarly to ordinary photon-subtracted and photon-added squeezed
vacuum states \cite{Dakna1,Dakna2}, the PSJP and PAJP squeezed vacuum 
states are Schr\"{o}dinger-cat-like states. From Eq.~(\ref{2.9}) the 
normalization constant ${\cal N}_{n,m}$ can be calculated to be
\begin{eqnarray}
{\cal N}_{n,m}=\sum_{k=\mu}^{n}\frac{(-|R|^2)^k}{(k-\nu)!}{n\choose k}
\sum_{j=\mu}^{n}\frac{(-|R|^2)^j}{(j-\nu)!}{n\choose j}
\sum_{p=\epsilon}^{\infty}
\frac{(2p+k)!(2p+j)!}{(p!)^2(2p+\nu)!4^p}|\kappa'|^{2p},
\label{2.10}
\end{eqnarray}
where $\epsilon$ $\!=$ $\!\max(0,[(1$ $\!-$ $\!\nu)/2])$, $[x]$ being the 
integer part of $x$. Rewriting Eq.~(\ref{2.10}) as
\begin{eqnarray}
{\cal N}_{n,m}=\frac{1}{|\kappa'|^\nu}\sum_{k=\mu}^{n}\frac{(-|R|^2)^k}
{(k-\nu)!}{n\choose k}
\sum_{j=\mu}^{n}\frac{(-|R|^2)^j}{(j-\nu)!}{n\choose j} 
\frac{\partial^{j-\nu}}{\partial |\kappa'|^{j-\nu}}|\kappa'|^j
\sum_{p=0}^{\infty}\frac{(2p+k)!}{(p!)^24^p}|\kappa'|^{2p}
\label{2.11}
\end{eqnarray}
and using the doubling formula for the gamma function 
$(2p+k)!$ $\!=$ $\!2^{2p+k}$ $\!\Gamma(p+\frac{k+1}{2})$ 
$\!\Gamma(p+\frac{k+2}{2})/\sqrt{\pi}$ and the definition of 
the hypergeometric function, we derive
\begin{eqnarray}
{\cal N}_{n,m}\!=\!\frac{1}{|\kappa'|^\nu}\sum_{k=\mu}^{n}\!\frac{k!(-|R|^2)^k}
{(k\!-\!\nu)!}{n\choose k}\!
\sum_{j=\mu}^{n}\!\frac{(-|R|^2)^j}{(j\!-\!\nu)!}{n\choose j}\! 
\frac{\partial^{j-\nu}}{\partial |\kappa'|^{j\!-\!\nu}}|\kappa'|^j
{\rm F}(\textstyle{\frac{k+1}{2},\frac{k+2}{2}},1;|\kappa'|^2).
\label{2.12}
\end{eqnarray}
Note that the hypergeometric function in Eq.~(\ref{2.12}) is  
proportional to a Legendre polynomial, 
${\rm F}(\frac{k+1}{2},\frac{k+1}{2},1,|\kappa'|^2)$ $\!=$
$\!(1\!-\!|\kappa'|^2)^{-(k+1)/2}$ $\!\rm{P}_k
[(1\!-\!|\kappa'|)^{-1/2}]$, so that Eq.~(\ref{2.12}) may be  
given by
\begin{eqnarray}
\lefteqn{
{\cal N}_{n,m}=\sum_{k=\mu}^{n}\frac{(-|R|^2)^k}
{(k-\nu)!}{n\choose k}
\sum_{j=\mu}^{n}\frac{j!(-|R|^2)^j}{(j-\nu)!}{n\choose j}
\sum_{l=\delta}^{j-\nu}{j-\nu\choose l}\frac{(k+2l')!|\kappa'|^{2l'}}
{2^{2l'-l}(l+\nu)!l'!}
}
\nonumber \\ && \hspace{-4ex}\times
\frac{\Gamma(\textstyle{l'+1/2})}{\Gamma(\textstyle{2l'-l+1/2})}
\,_3{\rm F}_2(\textstyle{l'+\frac{k+1}{2},l'+\frac{k+2}{2},
l'+\frac{1}{2},l'+1,2l'-l+1/2};|\kappa'|^2),
\label{2.13}
\end{eqnarray} 
where $l'\!=\![(l+1)/2]$. The  hypergeometric function 
$_3{\rm F}_2(a_1,a_2,a_3,b_1,b_2;z)$ in Eq.~(\ref{2.13}) can be 
calculated using standard routines. In particular, it may be expressed 
in terms of the more familiar hypergeometric function 
$_2{\rm F}_1(a_1,a_2,b_1;z)$ $\!\equiv$ $\!{\rm F}(a_1,a_2,b_1;z)$. 
The probability $P(n,m)$ of producing the state $|\Psi_{n,m}\rangle$
is found from Eq.~(\ref{2.6}) to be
\begin{equation}
P(n,m)=\frac{m!}{n!}
\frac{(1-|\kappa|^2)^{1/2}}{|R|^{2\nu}|T|^{2m}}{\cal N}_{n,m} .
\end{equation}

To illustrate the properties of $|\Psi_{n,m}\rangle$, in
Fig.~2 the quadrature-component distributions and the
Wigner function of a PSJP squeezed vacuum state for $n$ $\!=$ $\!1$ 
and \mbox{$m$ $\!=$ $\!4$} are plotted [$P(n,m)$ $\!=$ $\!3.37\%$]. From
an inspection of the figure, the state is seen to exhibit all the 
typical features of a Schr\"odinger-cat-like state. In particular, 
a more detailed analysis reveals that the Wigner function is a 
superposition of two quasi-Gaussian lobes with an interference structure 
between them. Since the corresponding formulas are rather lengthy, here we  
do without them and only give an explicit expression of the Husimi function,      
$Q(x,y)$ $\!=$ $\!|\langle\alpha|\Psi_{n,m}\rangle|^2/(2\pi)$, 
$\alpha$ $\!=$ $\!2^{-1/2}(x+iy)$, which takes the form
\begin{eqnarray}
\lefteqn{
Q_n(x,y|m)
=\frac{|R|^{4\nu}|T|^{4m}}{2\pi{\cal N}_{n,m}}
|\alpha|^{2\nu}e^{-|\alpha|^2}\exp\!\left\{{\textstyle\frac{1}{2}}
({\kappa'}^\ast\alpha^2+\kappa'{\alpha^\ast}^2) \right \}
}
\nonumber \\ && \hspace{10ex} \times \,
\Bigg |\sum_{k=\delta}^m{n\choose k\!+\!\nu}
\left(
\frac{|R|^2}{|T|^2}\sqrt{{\textstyle\frac{1}{2}} \kappa'{^\ast} \alpha }
\right)^k
{\rm H}_k\!\left(\sqrt{{\textstyle\frac{1}{2}} \kappa'{^\ast} \alpha } 
\right)\Bigg|^2.
\end{eqnarray}

%%%%%%%%%%%%%%%%%%%%%%%%%%%%%%%%%%%%%%%%%%%%%%%%%%%%%%%%%%%%%%%%%%%%%%%

\section{$N$-fold photon cho\-pping versus single-detector 
\protect\linebreak photocounting}
\label{sec3}

Among the practical problems that may be encountered in an experimental
generation of the Schr\"odinger-cat-like states, the losses associated
with nonperfect photon-number measurement may be of primordial
importance. For the sake of transparency let us restrict attention to the 
simplest situation and assume that an ordinary vacuum enters the second 
input of the beam splitter \cite{Dakna1}. In this case Eq.~(\ref{2.9}) 
reduces to
\begin{equation}
|\Psi_{0,m}\rangle\equiv|\Psi_{m}\rangle={\cal N}_{m}^{-1/2}
\sum_{n=0}^{\infty}  \frac{{\rm H}_{m+n}(0)}{\sqrt{n!}}
\left(-{\textstyle\frac{1}{2}}\kappa'\right)^{(n+m)/2} |n\rangle,
\label{3.0}
\end{equation}
with  
\begin{equation}
\label{3.0b}
{\cal N}_m 
= \frac{i^m m!}{|\kappa'|}
\left(\frac{|\kappa'|^2}{1-|\kappa'|^2}\right)^{(m+1)/2}
{\rm P}_m\!\left(-i\sqrt{\frac{|\kappa'|^2}{1-|\kappa'|^2}}\right)
\end{equation}
[$P_n(z)$, Legendre Polynomial],\footnote{Note that the finite sum in 
   \cite{Dakna1} can be expressed in terms of $P_n(z)$, on using the relation
   $n!\sum_{k=0}^{[n/2]}(2|x|)^{-2k}(k!)^{-2}\{(n-2k)!\}^{-1}$ $\!=$
   $\!i^n(1-|x|^2)^{n/2}|x|^{-n}{\rm P}_n(-i|x|/\sqrt{1-|x|^2})$.} 
and the probability 
$P(0,m)$ $\!\equiv$ $\!P(m)$ of producing the state $|\Psi_m\rangle$
can be given by 
\begin{equation}
\label{3.0c}
P(m)=\frac{i^m|R|^{2m}|\kappa'|^m\sqrt{1-|\kappa|^2}}{|T|^{2m}
(1-|\kappa'|^2)^{(m+1)/2}} \,
{\rm P}_m\!\left(-i\sqrt{\frac{|\kappa'|^2}{1-|\kappa'|^2}}\right),
\end{equation}
which is nothing but the probability of detecting $m$ photons 
in the readout mode.

At present two types of highly efficient photodetectors are
available: the linear-response photodiodes suitable for measuring strong 
signals without single-photon resolution and the avalanche photodiodes that 
may achieve single-event discrimination but are then saturated.  
It has therefore been 
suggested to spread a field whose photon-number statistics is desired to be 
detected over an array of such highly efficient avalanche photodiodes, 
on using passive optical multiports. Following \cite{Paul1}, we refer to 
this method as photon-chopping and to the photodiodes as type $I$ detectors. 
For a $2N_I$-port apparatus the probability of recording $k$ coincident 
events when $m$ photons are present is given by       
\begin{equation}
\label{3.1}
\tilde P_{N_I}(k|m) = \frac{1}{N_I^{m}} {N_I \choose k}
\sum_{l=0}^{k}(-1)^{l} {k \choose l} (k - l)^{m}
\end{equation}
for $k \le m$, and $\tilde P_{N_I}(k|m)\!=\!0$ for $k\!>\!m$. Note that
$\tilde P_{N_I}(k|m) \!\to\! \delta_{k,m}$ for $N_I \!\to\! \infty$.
In Eq.~(\ref{3.1}) perfect detection is assumed. The
effect of nonperfect detection may be modelled by placing an absorber in front
of the signal before it enters the $2N_I$ port. This corresponds to a random 
process such that photons are excluded from detection with probability 
$1\!-\!\eta_I$, $\eta_I$ being the efficiency of the photodiodes 
(Note that typically $\eta_I$ $\!=$ $\!0.8\ldots0.94$).
The probability of recording $k$ coincident events then modifies to
\begin{equation}
\label{3.2}
\tilde P_{N_I,\eta_I}(k|m) = \sum_{l} \tilde P_{N_I}(k|l) \, M_{l,m}(\eta_I) ,
\end{equation}
where the matrix $M_{l,m}(\eta_I)$ is given by
\begin{equation}
\label{3.3}
{M}_{l,m}(\eta_I) = {m \choose l} \eta_I^{l} (1 - \eta_I )^{m-l}
\end{equation}
for $l\!\le\!m$, and ${M}_{l,m}(\eta_I)\!=\!0$ for $l\!>\!m$.
Since detection of $k$ coincident events can result from various numbers
$m$ of photons, the conditional measurement yields a statistical mixture 
\begin{eqnarray}
\label{3.4}
\hat{\varrho}_{I}(k)
= \sum_{m}  P_{N_I,\eta_I}(m|k) \,|\Psi_{m} \rangle\langle \Psi_{m} |
\end{eqnarray}
rather than a pure state $|\Psi_{m}\rangle$. In Eq.~(\ref{3.4}),
$P_{N_I,\eta_I}(m|k)$ is the probability of $m$ photons being present
under the condition that $k$ coincident events are recorded. The conditional
probability $P_{N_I,\eta_I}(m|k)$ can be obtained using the Bayes rule,
\begin{eqnarray}
\label{3.5}
P_{N_I,\eta_I}(m|k) = \frac{1}{\tilde P_{N_I,\eta_I}(k)} 
\tilde P_{N_I,\eta_I}(k|m) \, P(m).
\end{eqnarray}
Here, $P(m)$ is the prior probability (\ref{3.0c}) of
$m$ photons being present, and accordingly, $\tilde P_{N_I,\eta_I}(k)$
is the prior probability of recording $k$ coincident events,
\begin{eqnarray}
\label{3.6}
\tilde P_{N_I,\eta_I}(k) = \sum_{m} \tilde P_{N_I,\eta_I}(k|m)\ P(m) .
\end{eqnarray}
 
Let us now consider a so-called type $II$ detector that is
able to discriminate between zero, one and a few more photons, but with
low detection efficiency ($\eta_{II}$ $\!\sim$ $\!0.3$). 
Using such a (single) detector for measuring the photon 
number yields
\begin{eqnarray}
\label{3.7}
%\hat \varrho_{\rm out}(0,k)\equiv
%\hat \varrho_{\rm out}(k)_{II}
\hat \varrho_{II}(k)
= \sum_{m}  P_{\eta_{II}}(m|k) \,|\Psi_{m} \rangle\langle \Psi_{m} |,
\end{eqnarray}
where, according to the Bayes rule,
the conditional probability $P_{\eta_{II}}(m|k)$ is now given by
\begin{eqnarray}
\label{3.8}
P_{\eta_{II}}(m|k)=\frac{1}{P_{\eta_{II}}(k)}M_{k,m}(\eta_{II})P(m),
\end{eqnarray}
with
\begin{eqnarray}
\label{3.8a}
P_{\eta_{II}}(k) = \sum_{m} M_{k,m}(\eta_{II})P(m).
\end{eqnarray}

In order to compare the conditional output states 
that are produced in the two schemes of photon-number measurement, we have 
calculated the (dimensionless) Shannon entropy
\begin{eqnarray}
\label{3.9}
S_{I(II)} = - \sum_{m} P_{I(II)}(m|k) \,\ln P_{I(II)}(m|k)
\end{eqnarray}
of the statistical mixtures of states,
\begin{eqnarray}
\label{3.9a}
\hat{\varrho}_{I(II)}(k) = \sum_{m} P_{I(II)}(m|k) \, 
|\Psi_m\rangle\langle\Psi_m| ,
\end{eqnarray}
as given by Eqs.~(\ref{3.4}) and (\ref{3.7}). The Shannon entropy is 
a measure of the spread of the distribution $P_{I(II)}(m|k)$, i.e., it is a 
measure of the deviation of $\hat{\varrho}_{I(II)}$ from a pure state. 
Note that for a pure state $S_{I(II)}$ $\!=$ $\!0$ is valid. From Fig.~3 we 
see that $S_I$ can always be reduced below $S_{II}$
when the number of type $I$ detectors in the photon chopping
scheme is sufficiently increased. Hence photon chopping
(with type $I$ detectors) may be much more suitable for     
preserving the quantum interference features in the 
conditional output state than the use of a single type $II$ 
detector. Photon chopping yields a mixed state that
is less spread and ``more pure'' than the state obtained
by using a single type $II$ detector for photon-number measurement.
Clearly, when in the photon chopping scheme type $II$ detectors 
are used in order to record coincident events, then
this scheme is less suitable for photon counting than a single
type $II$ detector. For $\eta_I$ $\!=$ $\!\eta_{II}$ the two schemes yield equal 
conditional output states only in the limit when $N_I$ $\!\to$ $\!\infty$.
For finite $N_I$ there is always a nonvanishing probability that
the number of recorded coincident events is smaller than the
number of photons, Eq.~(\ref{3.1}). In particular with regard to 
pure-state generation we have
\begin{eqnarray}
\lim_{N_{I}\to\infty} \lim_{\eta_{I}\to 1} S_{I} = 0,
\quad
\lim_{\eta_{II}\to 1} S_{II} = 0 .
\label{3.9b}
\end{eqnarray}
 
The conditional output states (\ref{3.4}) and (\ref{3.7})
can be determined using balanced homodyne detection and
measuring the quadrature-component distributions
\begin{eqnarray}
p_{I(II)}(x,\varphi|k) = \sum_m P_{I(II)}(m|k) \, p(x,\varphi|m),
\label{3.10a}
\end{eqnarray}
where \cite {Dakna1} 
\begin{eqnarray}
p(x,\varphi|m) = |\langle x,\varphi|\Psi_m\rangle|^2
= \frac{|\kappa' |^{m}}{2^{m}{\cal N}_m \sqrt{\pi \Delta^{m+1}}}
\exp\!\left( - \frac{1-|\kappa'| ^2}{\Delta} \, x^{2} \right)
\left |{\rm H}_m\!\left(K x\right)\right |^2 ,
\label{3.10}
\end{eqnarray}
with $\Delta$ $\!=$ $\!1$ $\!+$ $\!|\kappa'|^2$ $\!+$ 
$\!2 |\kappa'|\cos(2\varphi-\varphi_{\kappa'})$ and 
$K$ $\!=$ $\!\sqrt{(-{\kappa'}^\ast e^{i2\varphi}-|\kappa'|^2)/\Delta}\,$. 
In Fig.~4 we report the results of  simulated measurements. In the case of 
a single type $II$ detector, Fig.~4(a),
the quantum interferences are totally smeared and non-observable.
Although somewhat smeared, the quantum interferences are 
observable in a photon-chopping scheme with type $I$ detectors,
Fig.~4(b). The result obviously reflects the above mentioned
fact that the conditional output state $\hat{\varrho}_{II}$,
Eq.~(\ref{3.7}), is ``more mixed'' than the state
$\hat{\varrho}_{I}$, Eq.~(\ref{3.4}), in general.   

The reconstruction of the states $|\Psi_{m}\rangle$ from the 
mixed state $\hat \varrho_{II}(k)$, Eq.~(\ref{3.7}), can be achieved 
using the inverse Bernoulli transform \cite{DARIANO1},
\begin{eqnarray} 
|\Psi_{m} \rangle\langle \Psi_{m} |
=\frac{1}{P(m)\eta_{II}^m}
\sum_{k=m}^{\infty}{k\choose m}(1\!-\!\eta_{II}^{-1})^{k-m}
P_{\eta_{II}}(j) \hat \varrho_{II}(k).
\label{3.11}
\end{eqnarray}
Similarly, inverting Eq.~(\ref{3.4}), we obtain  ($N_I$ $\!\to$ $\! \infty$)
\begin{eqnarray}
\label{3.12}
|\Psi_{m} \rangle\langle \Psi_{m} |
= \frac{1}{P(m)\eta_{I}^m}
\!\sum_{j=m}^{\infty}\!{j\choose m}(1\!-\!\eta_{I}^{-1})^{j-m}
\left[\sum_{k=j}^{\infty} (\tilde P_{N_I})^{-1}_{j,k}
\tilde P_{N_I,\eta_I}(k) \hat \varrho_{I}(k) \right],
\end{eqnarray}
where  $(\tilde P_{N_I})^{-1}_{m,k}$ is the inverse
of the matrix $(\tilde P_{N_I})_{k,m}$
$\equiv$ $\!\tilde P_{N_I,\eta_I}(k|m)$, and the following
recursion relation is valid \cite{Paul1}: 
\begin{eqnarray}
(\tilde P_{N_I})^{-1}_{n,n+k}
=\frac{-1}{(\tilde P_{N_I})_{n+k,n+k}}
\sum_{j=0}^{k-1}
(\tilde P_{N_I})^{-1}_{n,n+j}
(\tilde P_{N_I})_{n+j,n+k},\quad k=0,1,2\cdots \,.
\label{3.13}
\end{eqnarray}
Note that Eq.~(\ref{3.12}) reproduces the exact components 
$|\Psi_{m} \rangle\langle \Psi_{m} |$ only for $N_I \!\to\! \infty$.

Applying Eq.~(\ref{3.11}) and Eq.~(\ref{3.12}) (for finite $N_I$),
the states $|\Psi_m\rangle$ can be reconstructed from the 
homodyne data of the measured mixed states. Examples of reconstructed 
quadrature-component distributions are shown in Fig.~5. From the 
figure we see that processing the homodyne data according to
Eqs.~(\ref{3.11}) and (\ref{3.12}) allows us to restore the quantum 
interferences in the quadrature-component distributions of the
component states of the produced statistical mixtures of states. 
It should be noted that with increasing number of measurements 
the Bernoulli inversion in Eq.~(\ref{3.11}) yields the almost
perfect interference structure [Fig.~5(a)]. 
In order to realize (for the same number of measurements) a comparable 
accuracy on the basis of Eq.~(\ref{3.12}), a sufficiently large number 
$N_I$ of channels (type $I$detectors) in the photon-chopping scheme must 
be used [$N_I$ $\!=$ $\!50$ in place of $N_I$ $\!=$ $\!20$ used
in Fig.~5(b)]. This is obviously due to the fact that 
in the photon-chopping scheme the probability that a photon impinges on an 
already saturated photodiode approaches zero only for $N_I$ $\!\to$ 
$\!\infty$. On the other hand, photon chopping already yields reasonable 
results for small amounts of data, even when the number of channels is
reduced [Fig.~5(d)]. From Fig.~5(c) it is seen that the error in the 
single-(type-$II$-)detector scheme drastically increases with decreasing 
number of measurements. This result tells us that photon-chopping may be
more powerful than a single-detector scheme when the produced
state tends to a macroscopical one and the amount of data needed
becomes great.

Finally, let us mention that with respect to the probability
of producing particularly macroscopic Schr\"odinger-cat-like states the 
chopping scheme may be more suitable than the single-detector
scheme. In fact, from  Fig.~6 we see that the probability of recording 
$k$ $\!>$ $\!0$ clicks is always higher for the photon-chopping method 
as for the single-detector scheme. This effect can be made a bit more 
pronounced choosing larger $N_I$.

Let us briefly comment on the use of a nonperfect homodyne detector 
for measuring the quadrature-component distributions (\ref{3.10a}).
Since in a realistic homodyne experiment the quadrature components cannot  
be measured exactly, we may assume that instead of 
$p_{I(II)}(x,\varphi|k)$ smeared distributions 
\begin{eqnarray}
p_{I(II)}(x,\varphi;\eta|k) 
= \sum_m P_{I(II)}(m|k) \, p(x,\varphi;\eta|m)
\label{3.14a}
\end{eqnarray}
are measured, where
\begin{eqnarray}
\label{3.14}
p(x,\varphi;\eta|m) 
= \int_{-\infty}^{\infty} dy \, f(x-y;\eta) \, p(y,\varphi|m) , 
\end{eqnarray}
$f(x;\eta)$ being some positive single-peaked function of $x$,
such as a Gaussian,
\begin{eqnarray}
\label{3.15}
f(x;\eta)=  \frac{1}{\sqrt{2\pi\sigma^2}} 
\exp \!\left( - \frac{x^2}{2\sigma^2}\right),
\quad \sigma = \frac{1-\eta}{2\eta} 
\end{eqnarray}
($\eta$, quantum efficiency of the homodyne detector, with
$\eta$ $\!\le$ $\!1$). Combining Eqs.~(\ref{3.14}), (\ref{3.10}),
and (\ref{3.15}) and performing the $y$-integration yields
\begin{eqnarray}
\label{3.16}
\lefteqn{
p(x,\varphi;\eta|m) 
=\frac{|\kappa' |^{m}|K|^{2m}}{p^{m+1/2}
{\cal N}_m \sqrt{2\pi \Delta^{m+1}\sigma^2}}
\exp\!\left( - \frac{1-(2p\sigma^2)^{-1}}{2\sigma^2} \, x^{2} \right)
}
\nonumber \\ && \hspace{10ex}\times
\sum_{k=0}^{m}{m\choose k}^2(m-k)!
\left(\frac{|p-K^2|}{2|K|^2}\right)^k
\left|{\rm H}_k\!\left(\frac{K x}
{2\sigma^2\sqrt{p^2-p K^2}}\right)\right|^2 \!,
\end{eqnarray}
$p$ $\!=$ $\!(1$ $\!-$ $\!|\kappa'|^2)/\Delta$ $\!+$ $\!1/(2\sigma)$. 
It can be readily seen from Eqs.~(\ref{3.14}) -- (\ref{3.16}) that 
$p(x,\varphi;\eta|m)$ $\!\to$ $\!p(x,\varphi|m)$ for 
$\eta$ $\!\to$ $\!1$. The notorious fragility of the interference 
structure may be directly seen by comparing the quadrature-component
distributions (\ref{3.10}) and (\ref{3.16}). For example, for
$m$ $\!=$ $\!3$ (an a mean number of photons  
$\langle\hat n\rangle$ $\!=$ $\!15$) the interference structure 
is completely smeared out if $\eta$ $\!<$ $\!0.94$ (Fig.~7). 

It is well known that for compensating the losses of the homodyne 
detector one can make a detour via the density matrix in the Fock 
representation. The density-matrix elements are reconstructed from 
the noisy quadrature-component distributions using loss-compensating 
kernels,
\begin{eqnarray}
\label{3.17}
\hat\varrho_{n,n'}(m)
=\langle n|\Psi_m\rangle\langle\Psi_m|n'\rangle
=\int_{2\pi}\,d\varphi
\int_{-\infty}^{\infty}\,d\,x\,p(x,\varphi;\eta|m)K_{n,n'}(x,\varphi;\eta),
\end{eqnarray} 
$K_{n,n'}(x,\varphi;\eta)$ being given in \cite{DARIANO2}
(note that such a compensation is possible only for $\eta$ $\!>$ $\!0.5$).
Alternatively one can first reconstruct the density-matrix elements in the 
Fock basis using the kernel functions for perfect detection and then 
apply an inverse Bernoulli transform to reconstruct the true
density-matrix elements \cite{Kiss1}.

\section{Summary}
\label{sec4}

We have shown that quantum-state preparation
via conditional output measurement on a beam splitter can be 
advantageously used for preparation of a great variety of 
Schr\"odinger-cat-like states. When a mode prepared in a squeezed vacuum 
and a mode prepared in an arbitrary 
Fock state are superimposed by a beam splitter
and an arbitrarily chosen number of photons is recorded
in one of the output channels of the beam splitter,
then the mode in the other output channel is prepared in either 
a photon-subtracted or a photon-added 
Jacobi polynomial squeezed vacuum state,
the latter being obtained by applying an operator-valued Jacobi
polynomial to a squeezed vacuum state.
All the PSJP and PAJP states represent examples of Schr\"odinger-cat-like
states, provided that the sum of the number of incident photons and the
number of detected photons is nonzero.

In order to project onto Fock states, we have studied
two methods of direct photon counting which may be realized
with currently available techniques: single-detector photon 
counting and $N$-fold photon chopping. Both methods produce
statistical mixtures of Schr\"odinger-cat-like states rather
than pure states in general, because of nonperfect detection.
We found that photon chopping offers the possibility of direct 
observation of the quantum interferences. Moreover this method 
can be advantageously used to reduce the amount of data needed 
for reconstructing the Schr\"odinger-cat-like states in the 
produced mixed state, which can be measured by balanced homodyning. 
When the number of recorded data is suitably large, then the use of 
a single (low-efficiency) detector may be more advantageous. In this 
case the pure-state components of the statistical mixture of states 
can simply be calculated from the measured data using the 
inverse Bernoulli transform. To realize the same accuracy
in photon chopping, the number of channels and detectors must 
be relatively high.

\noindent {\bf Acknowledgements}
This work was supported  by the Deutsche Forschungsgemeinschaft.

\newpage
\begin{figure}[tbh]
\centering\epsfig{figure=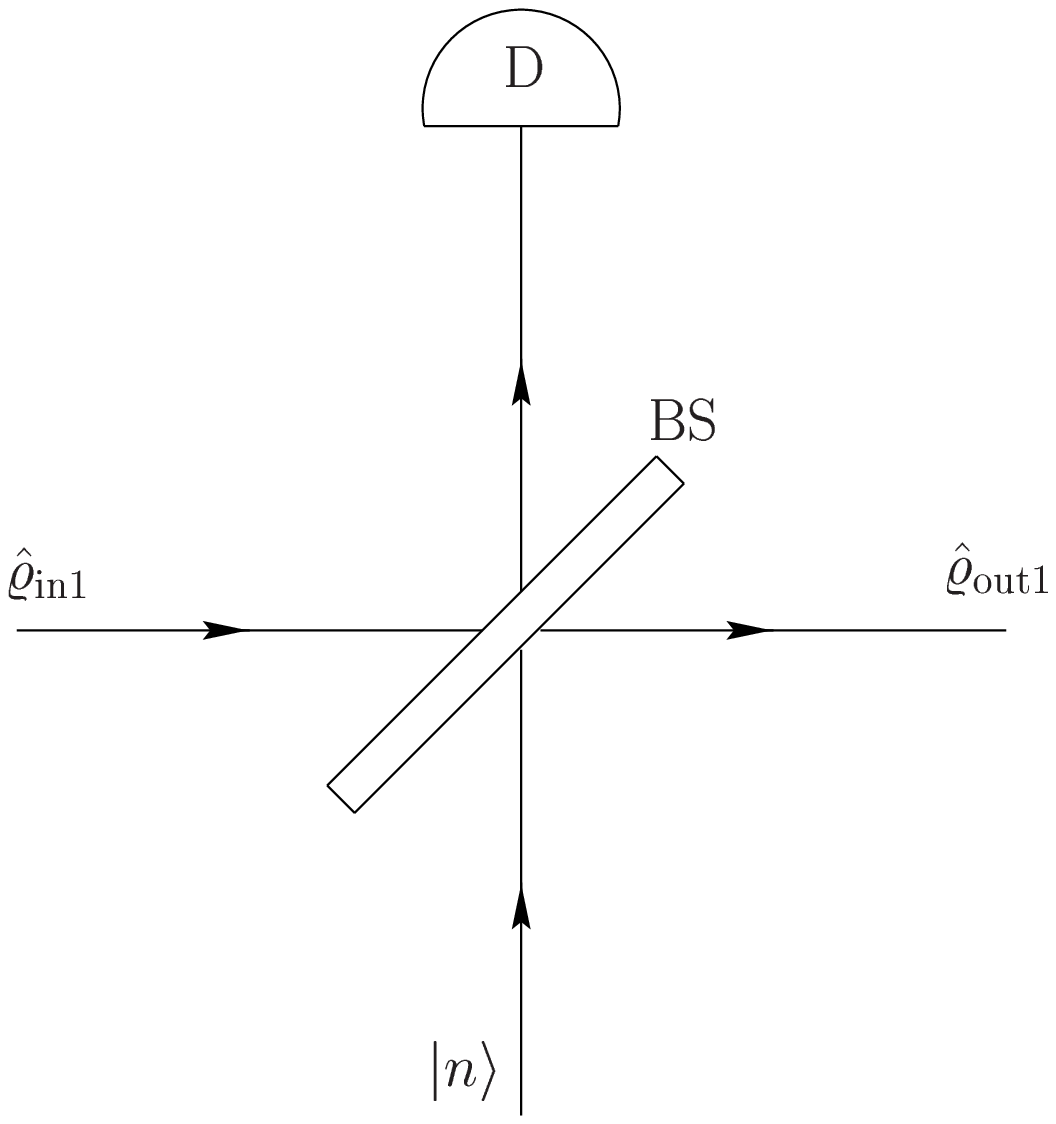,width=0.7\linewidth}
\caption{Experimental setup for generating the 
Schr\"odinger-cat-like states. A signal mode prepared in a state 
$\hat{\varrho}_{{\rm in}1}$ $\!=$ $\!\hat{S}(\xi)| 0 \rangle\langle 0  
| \hat{S}^{\dagger}(\xi)$ is mixed at the beam splitter BS with 
another input mode prepared in a Fock state $|n\rangle$, and $m$ photons 
are recorded by the detector D in one of the output channels of the 
beam splitter. The quantum state $\hat{\varrho}_{{\rm out}1}$ of the mode in the other output 
channel is found to be  ``collapsed'' to either a PSJP 
($n$ $\!<$ $\!m$) or a PAJP ($n$ $\!>$ $\!m$) squeezed vacuum state.} 
\end{figure}

\newpage
\begin{figure}[tbh]
\vspace{2cm}
\begin{minipage}[b]{0.45\linewidth}
\centering\epsfig{figure=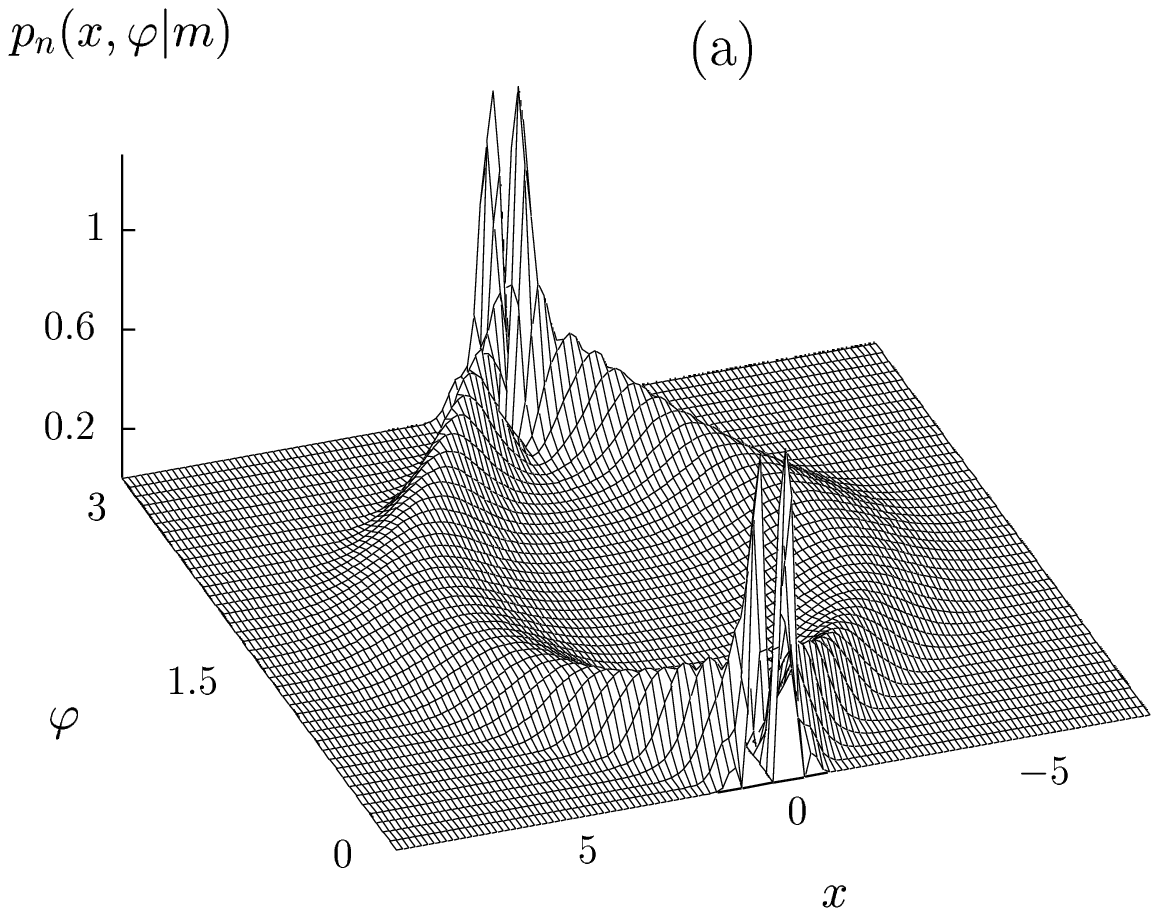,width=0.9\linewidth}
\end{minipage}\hfill 
\begin{minipage}[b]{0.45\linewidth}
\centering\epsfig{figure=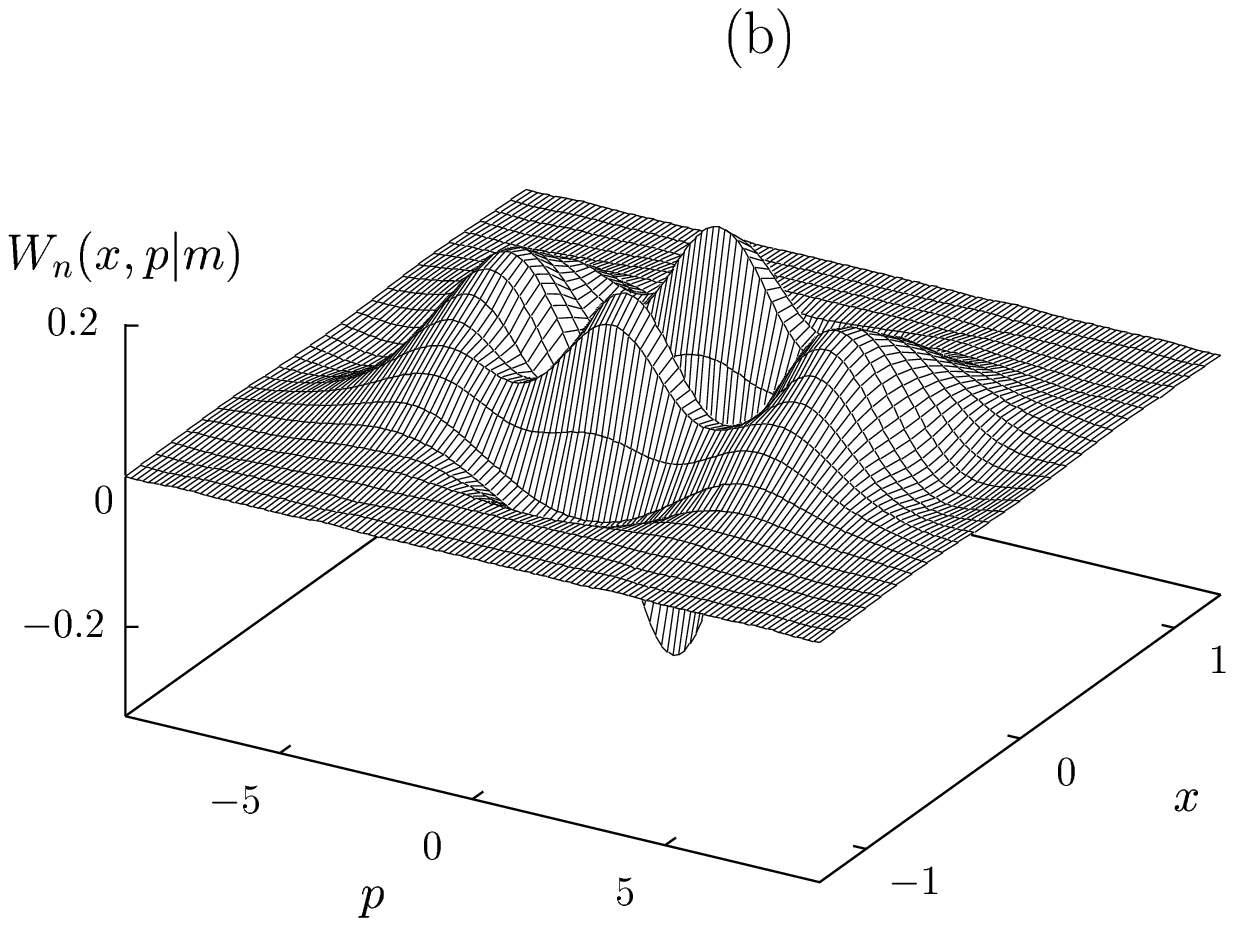,width=0.9\linewidth}
\end{minipage} 
\caption{Quadrature distribution (a) and the Wigner function (b) of a 
PSJP squeezed vacuum $|\Psi_{n,m}\rangle$ [$n\!=\!1$ and $m\!=\!4$] for
$\kappa'\!=-\!0.81$ [$|T|^2\!=\!0.9$, $|\kappa|\!=\!0.9$].}
\end{figure}
\newpage
\begin{figure}[tbh]
\vspace{2cm}
\begin{minipage}[b]{0.45\linewidth}
\centering\epsfig{figure=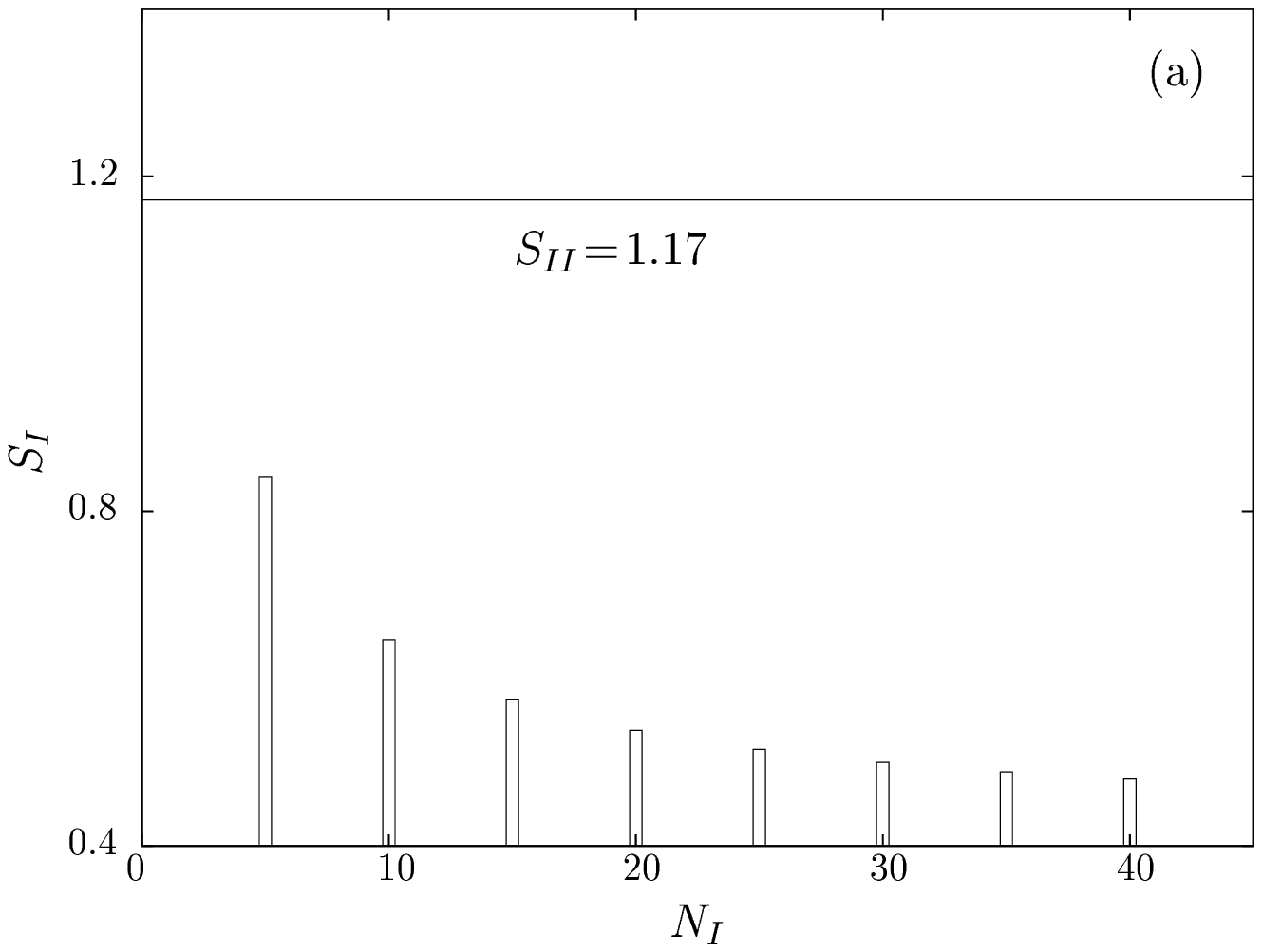,width=0.9\linewidth}
\end{minipage} \hfill
\begin{minipage}[b]{0.45\linewidth}
\centering\epsfig{figure=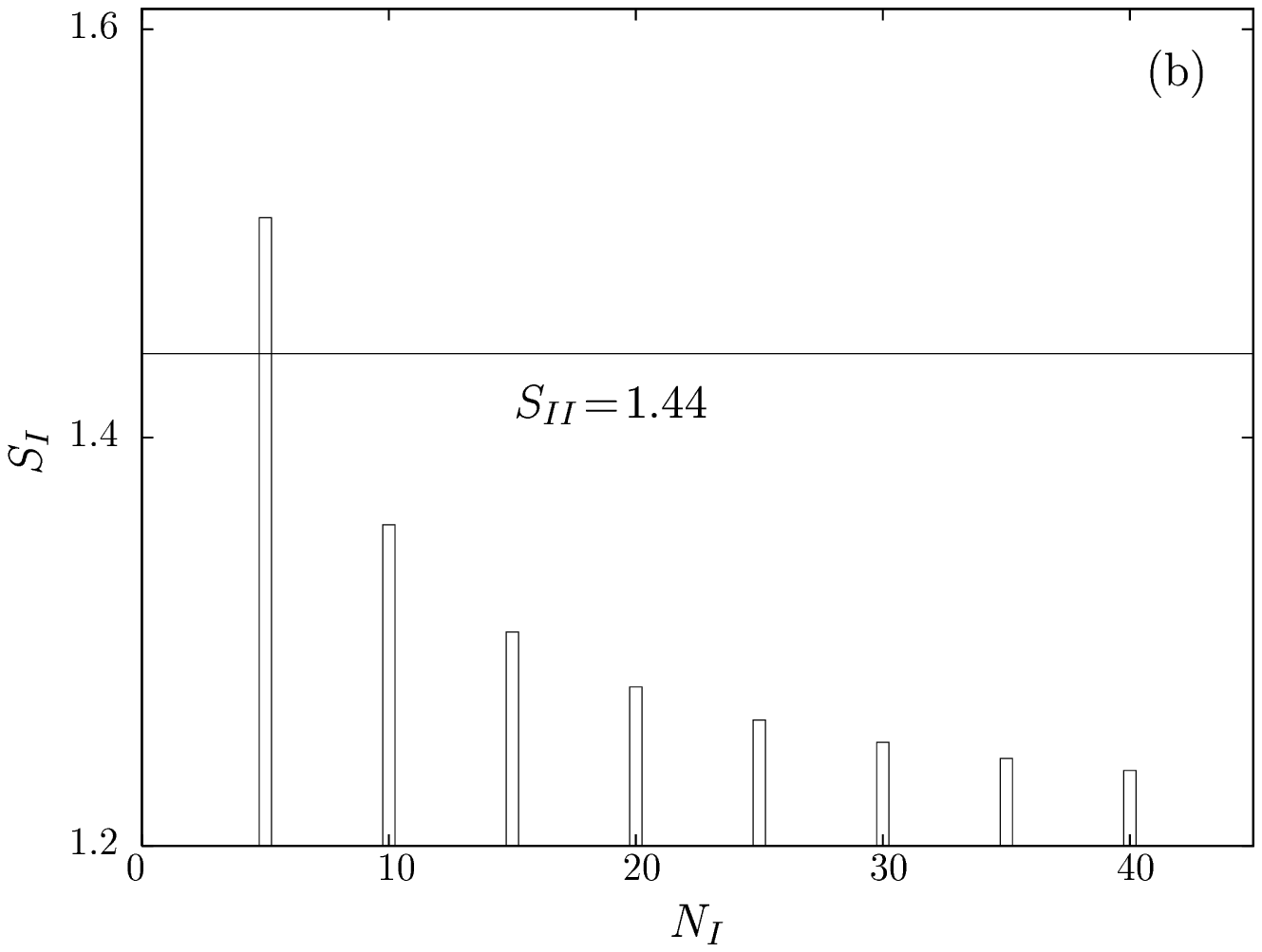,width=0.9\linewidth}
\end{minipage}
\caption{The Shannon entropies $S_I$ and $S_{II}$ of the mixed
states (\protect\ref{3.4}) and (\protect\ref{3.7}), respectively. In (a) it is
assumed that $k$ $\!=$ $\!3$ events are recorded and the  
quantum efficiencies are $\eta_I$ $\!=$ $\!0.85$ and
$\eta_{II}$ $\!=$ $\!0.3$. In (b) the parameters are
$k$ $\!=$ $\!5$ and $\eta_I\!=\!0.5$ and $\eta_{II}\!=\!0.3$.
The calculations are performed for $\kappa'$ $\!=$ $\!-0.7$ 
($|T|^2$ $\!=$ $\!0.9$, $|\kappa|$ $\!=$ $\!0.77$).}
\end{figure}

\newpage
\begin{figure}[tbh]
\vspace{2cm}
\begin{minipage}[b]{0.45\linewidth}
\centering\epsfig{figure=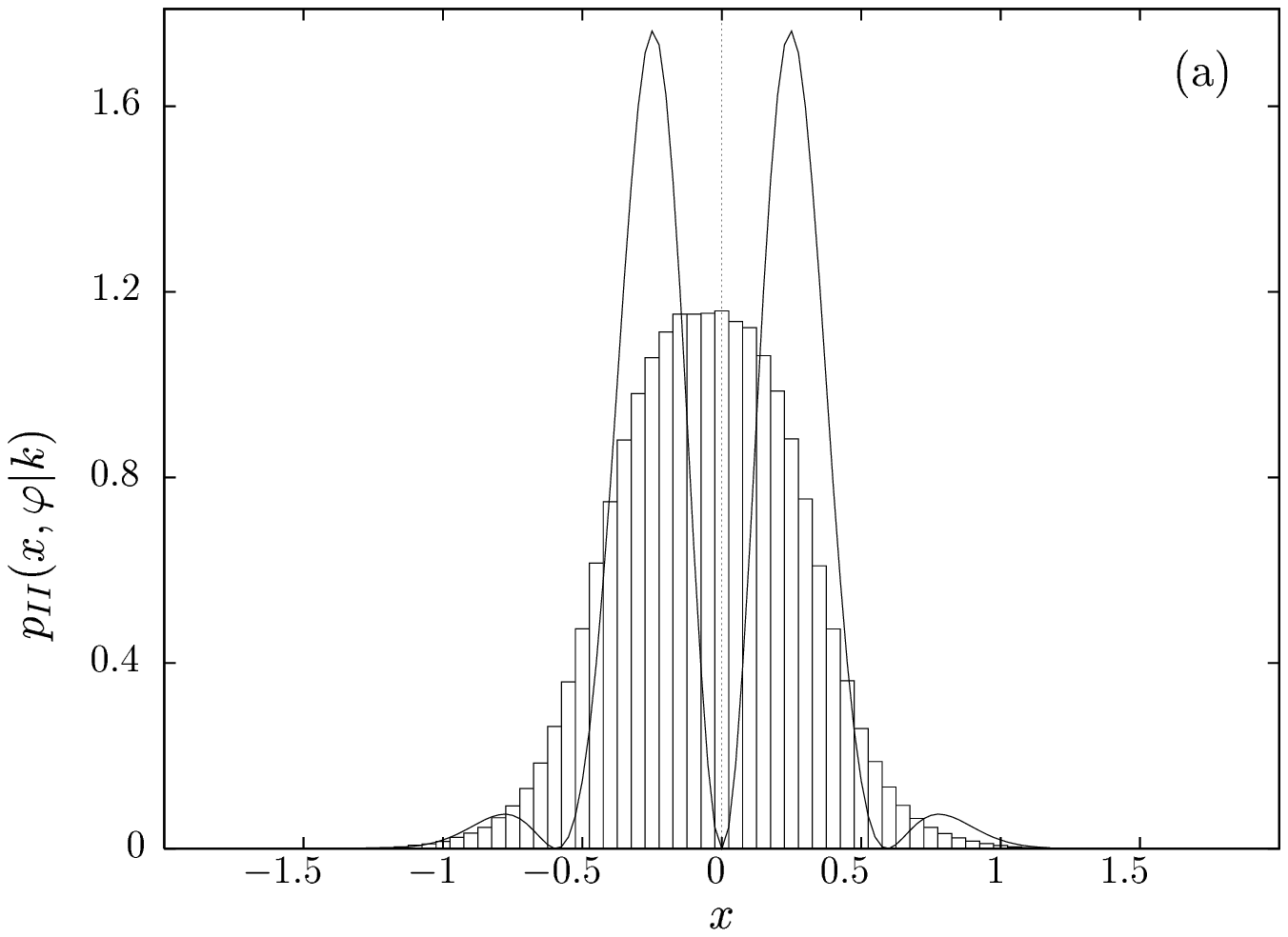,width=\linewidth}
\end{minipage}   \hspace{.5ex} 
\begin{minipage}[b]{0.45\linewidth}
\centering\epsfig{figure=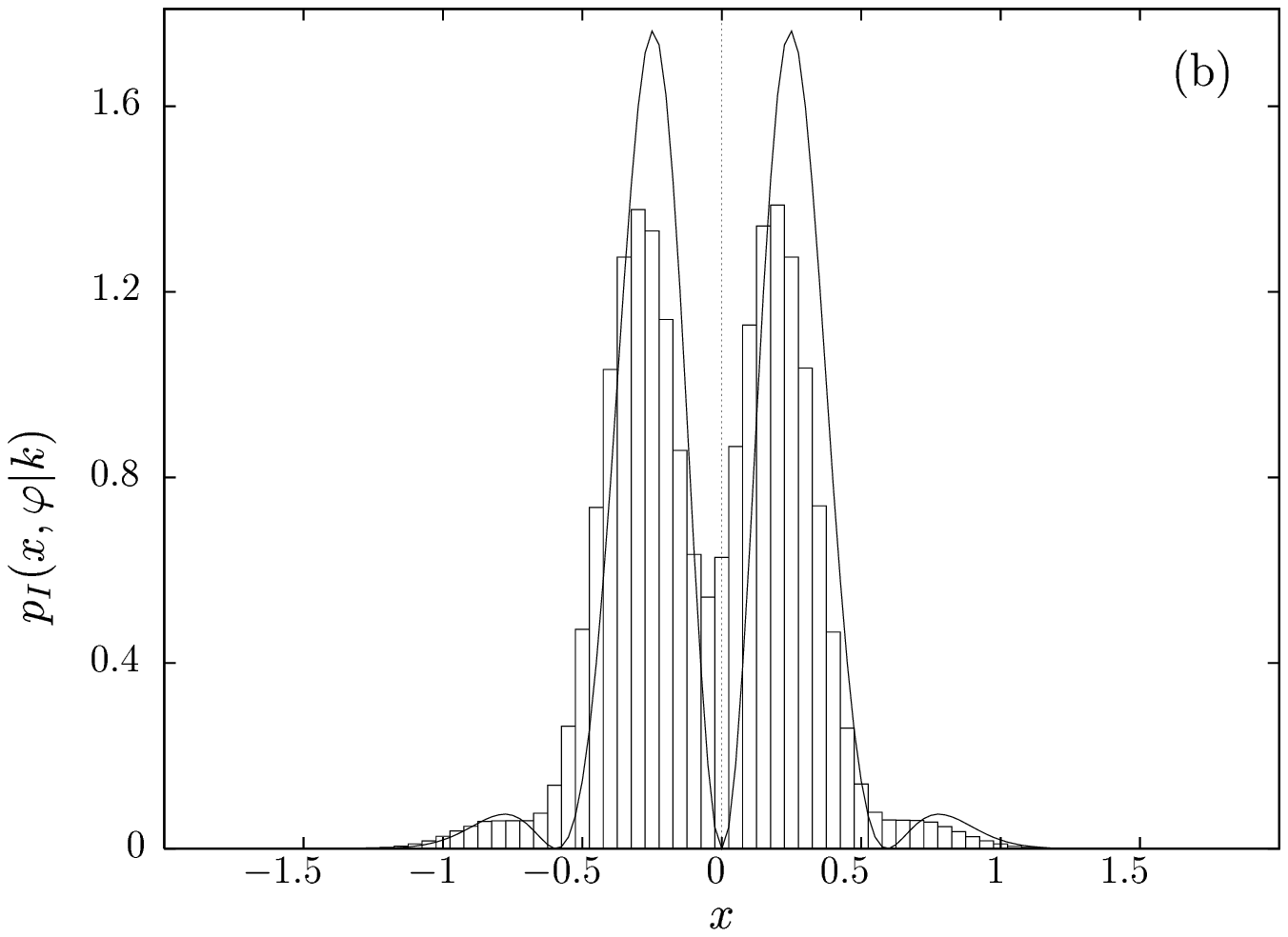,width=\linewidth}
\end{minipage}
\caption{ The quadrature-component distributions 
$p_{II}(x,\varphi|k)$ (a) and $p_{I}(x,\varphi|k)$ (b),
respectively, of the mixed states 
(\protect\ref{3.7}) ($\eta_{II}$ $\!=$ $\!0.3$) and 
(\protect\ref{3.4}) ($N_I$ $\!=$ $\!20$, $\eta_I$ $\!=$ $\!0.9$)  
are shown for $\varphi$ $\!=$ $\!0$ and $k$ $\!=$ $\!3$. 
The theoretical distributions (\protect\ref{3.10}) (solid lines)
are compared with the histograms obtained from computer-simulations
of $10^6$ measurements. 
The calculations are performed for $\kappa'$ $\!=$ $\!-0.81$ 
($|T|^2$ $\!=$ $\!0.9$, $|\kappa|$ $\!=$ $\!0.9$).}
\end{figure}
\newpage

\begin{figure}[tbh]
\vspace{2cm}
\begin{minipage}[b]{0.45\linewidth}
\centering\epsfig{figure=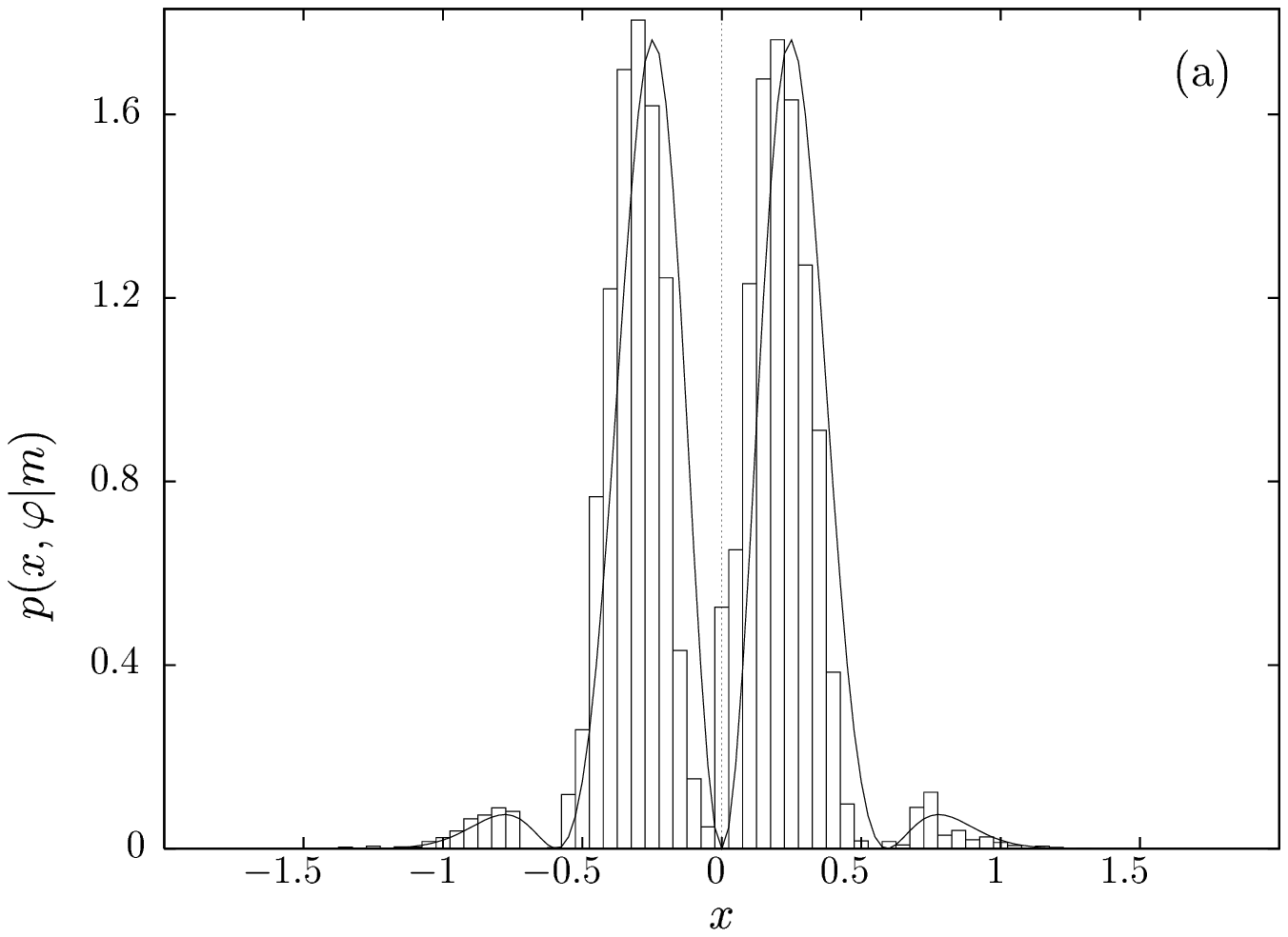,width=\linewidth}
\end{minipage} \hspace{.5ex}
\begin{minipage}[b]{0.45\linewidth}
\centering\epsfig{figure=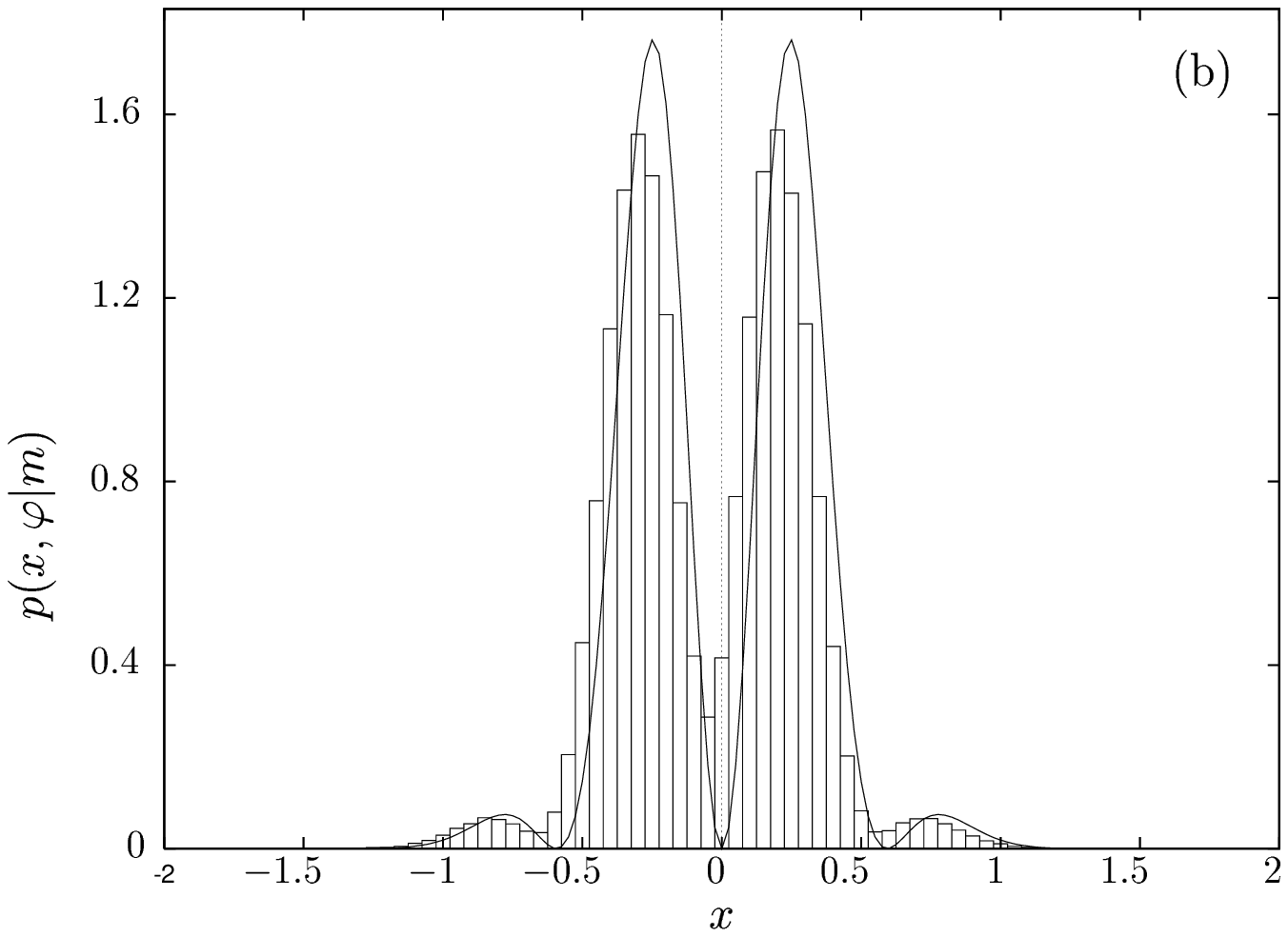,width=\linewidth}
\end{minipage}
\vspace{0.7cm}

\begin{minipage}[b]{0.45\linewidth}
\centering\epsfig{figure=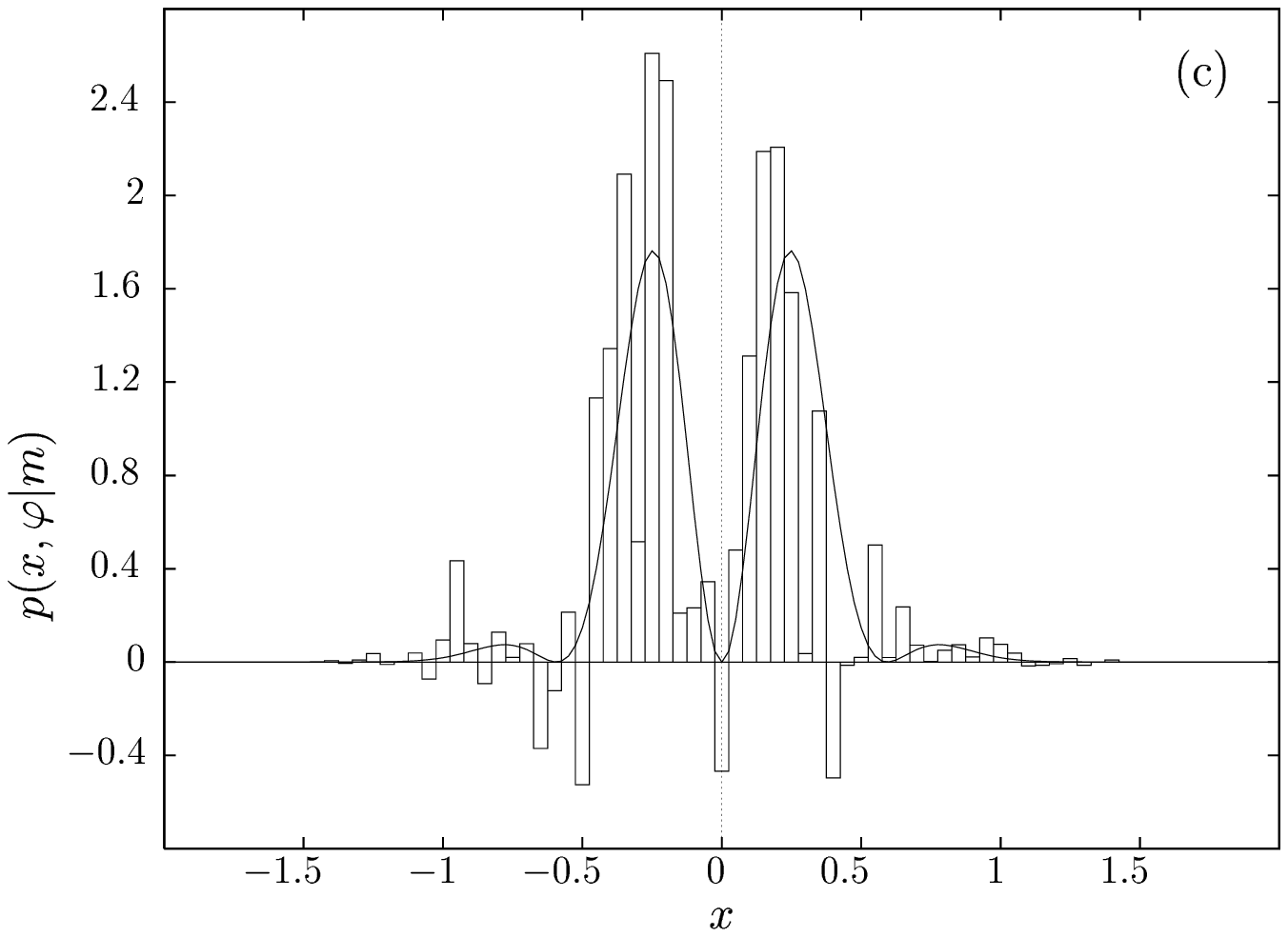,width=\linewidth}
\end{minipage}     \hspace{.5ex}
\begin{minipage}[b]{0.45\linewidth}
\centering\epsfig{figure=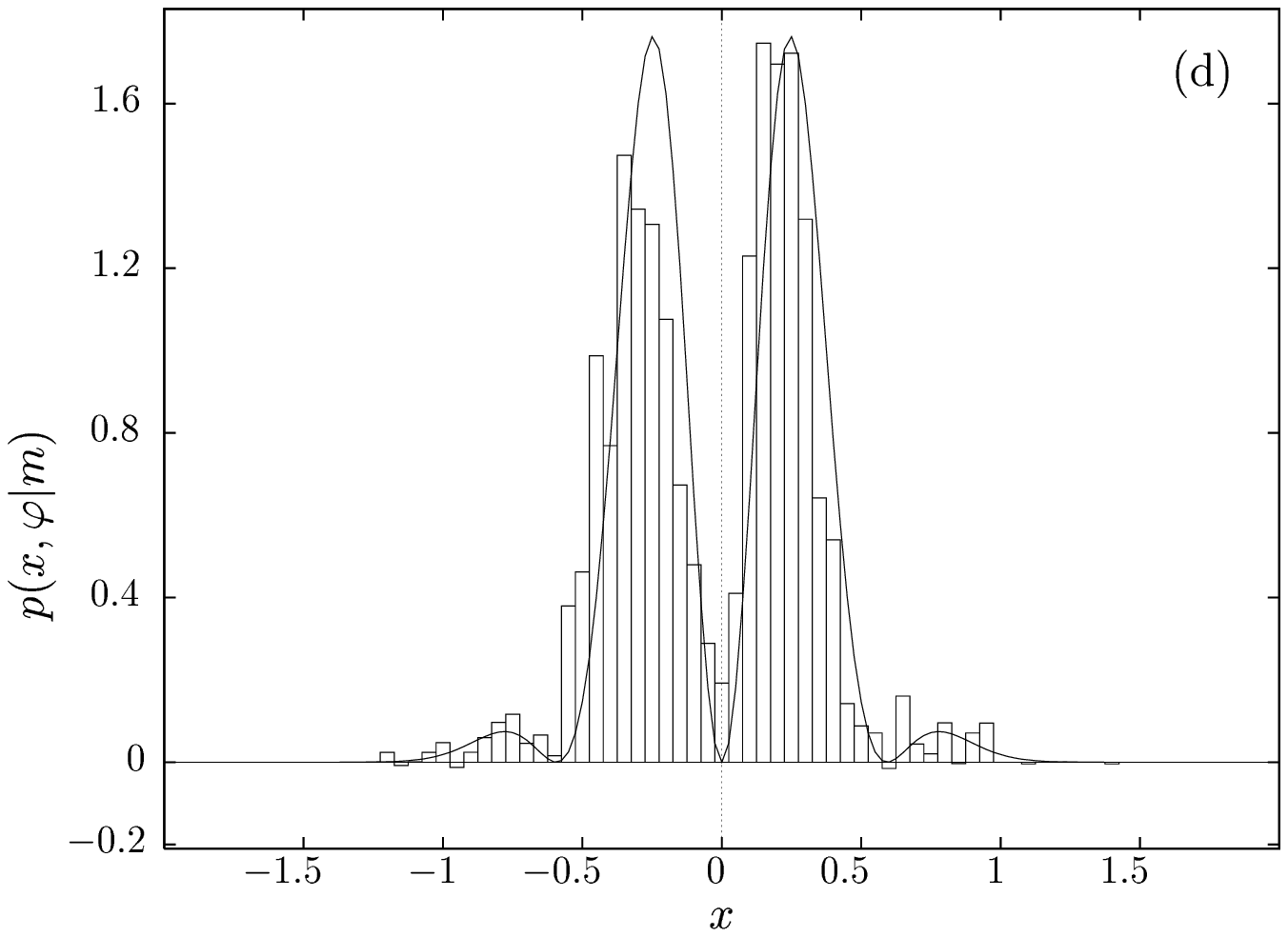,width=\linewidth}
\end{minipage}
\caption{The quadrature-component distributions 
$p(x,\varphi|m)$ of $|\Psi_m\rangle$
reconstructed from the homodyne data,
(a,c) according to Eq.~(\protect\ref{3.11}) ($\eta_{II}$ $\!=$ $\!0.3$)  
and (b,d) according to Eq.~(\protect\ref{3.12}) ($N_I$ $\!=$ $\!20$, 
$\eta_I$ $\!=$ $\!0.9$), are shown for $\varphi$ $\!=$ $\!0$ 
and $m$ $\!=$ $\!3$. The theoretical distribution (\protect\ref{3.10}) 
(solid line) is compared with the distributions reconstructed 
from (a,b) $5\cdot10^5$ measurement, (c) $10^4$ measurements,
and (d) $10^3$ measurements. 
The calculations are performed for $\kappa'$ $\!=$ $\!-0.81$ 
($|T|^2$ $\!=$ $\!0.9$, $|\kappa|$ $\!=$ $\!0.9$).}
\end{figure}

\newpage
\begin{figure}[tbh]
\hspace{-6ex}
\centering\epsfig{figure=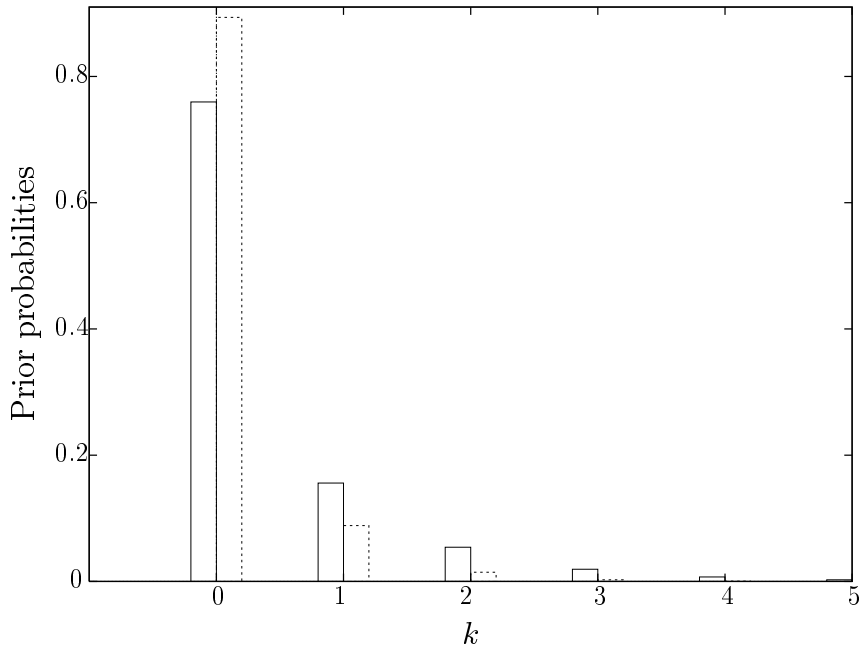,width=0.9\linewidth}
\caption{The prior probabilities 
(\protect\ref{3.6}) (solid bars) and (\protect\ref{3.8a}) (dashed bars)
for recording $k$ clicks in the photon-chopping scheme
($\eta_I$ $\!=$ $\!0.9$, $N_I$ $\!=$ $\!20$) and 
the single-detector scheme ($\eta_{II}$ $\!=$ $\!0.3$),
respectively, are shown. The calculations are performed for 
$\kappa'$ $\!=$ $\!-0.81$ 
($|T|^2$ $\!=$ $\!0.9$, $|\kappa|$ $\!=$ $\!0.9$).
Note that for $k\!=\!3$ the type $II$ detector
only gives $12\%$ of the probability obtained 
by photon chopping.}
\end{figure}

\newpage

\begin{figure}[tbh]
\hspace{-6ex}
\centering\epsfig{figure=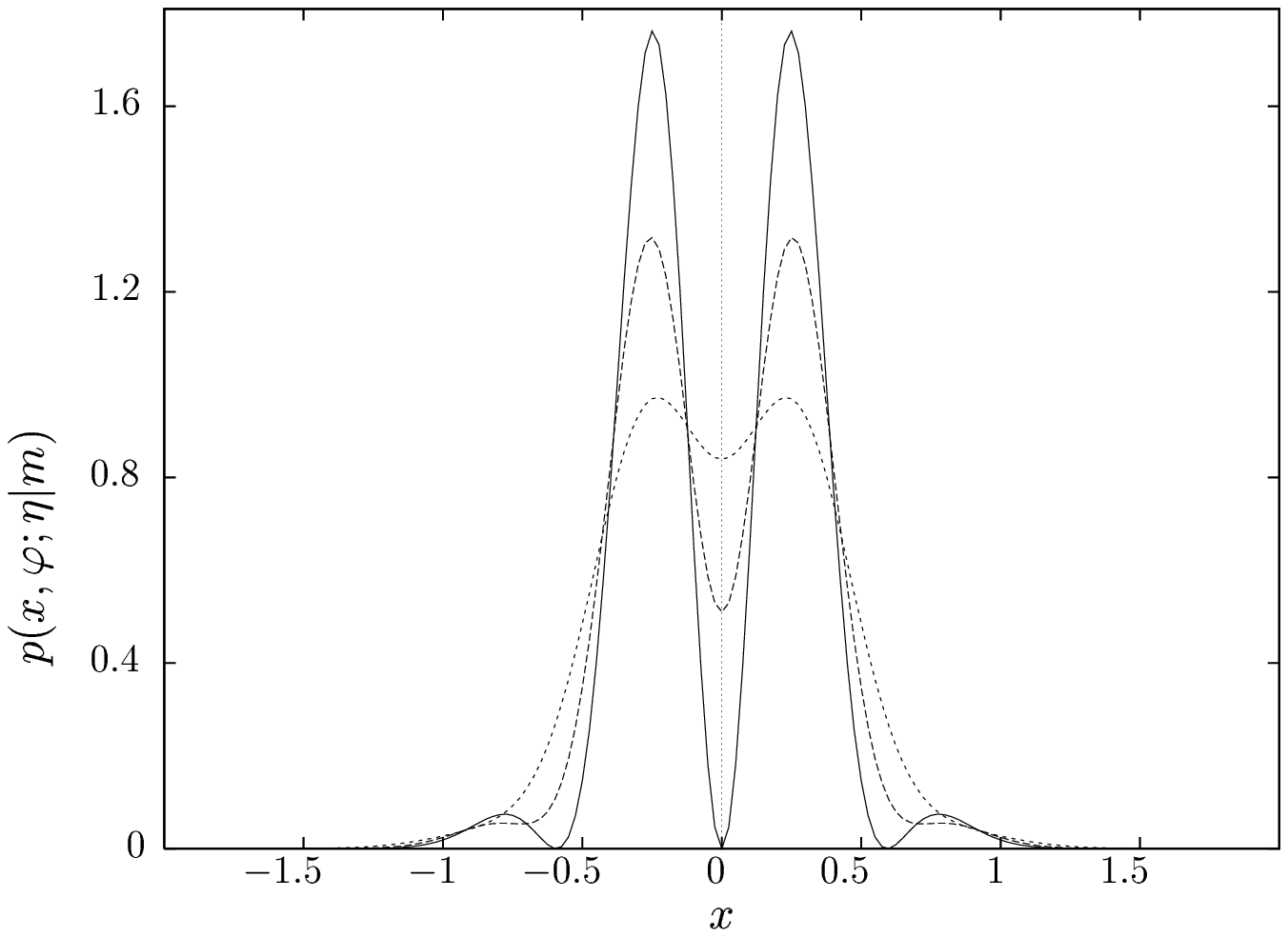,width=0.9\linewidth}
\caption{ The 
theoretical quadrature-component distribution 
(\protect\ref{3.10}) at the phase $\varphi$ $\!=$ $\!0$ (solid line) 
is compared with the quadrature-component distribution (\protect\ref{3.16}) 
obtained in nonperfect homodyne detection
for $\eta$ $\!=$ $\! 0.98$ (dashed line)  and 
$\eta$ $\!=$ $\! 0.94$ (dotted line). The values of the other 
parameters are as in Fig.~3
The calculations are performed for $\kappa'$ $\!=$ $\!-0.81$ 
($|T|^2$ $\!=$ $\!0.9$, $|\kappa|$ $\!=$ $\!0.9$).}
\end{figure}


\itemsep0pt
\begin{thebibliography}{99}

\bibitem{Schroedinger1}
E. Schr\"odinger, {Naturwissenschaften} {\bf 23}, {(1935)} {807}.

\bibitem{Zurek1}
W.H. Zurek, {Phys. Today} {\bf 44}, {(1991)} {36}; {\bf 46}, {(1993)} {81}.

\bibitem{Rauch}
D.L. Jacobson, S.A. Werner, H. Rauch.
{Phys. Rev. A} {\bf 49}, {(1994)} {3196}.

\bibitem{Monroe1}
C. Monroe, D.M. Meekhof, B.E. King, D.J. Wineland,
{Science} {\bf 272}, {(1996)} {1131}.

\bibitem{DARIANO1}
G.M. D'Ariano, C. Machiavello, L. Maccone, 
[{\sl Los Alamos e-print archive} quant-ph/9804021 (1998)]. 

\bibitem{Song1}
S. Song, C. M. Caves, B. Yurke,
{Phys. Rev. A} {\bf 41}, {(1990)} {5261}.

\bibitem{Yurke1}
B. Yurke, W. Schleich and D. F. Walls,
{Phys. Rev. A} {\bf 42}, {(1990)} {1703}.

\bibitem{Dakna1}
M. Dakna, T. Anhut, T. Opatrn\'{y}, L. Kn\"{o}ll, D.--G. Welsch,
{Phys. Rev. A} {\bf 55}, {(1997)} {3184}.

\bibitem{Dakna2}
M. Dakna, L. Kn\"{o}ll, D.--G. Welsch, to appear in 
{\sl Europ. Phys. J.  D} 
[{\sl Los Alamos e-print archive} quant-ph/9803077 (1998)].


\bibitem{Dakna3}
M. Dakna, L. Kn\"{o}ll, D.--G. Welsch {Opt. Commun.} {\bf 145}, {(1998)} {309}.

\bibitem{Paul1}
H. Paul, P. T\"orm\"a, T. Kiss, I. Jex,
{Phys. Rev. Lett.} {\bf 77}, (1996) 2446.

\bibitem{Campos1}
R.A. Campos, B.E.A. Saleh, M.C. Teich,
{Phys. Rev. A} {\bf 40}, {(1989)} {1371},



\bibitem{DARIANO2}
G. M. D'Ariano, U. Leonhardt, H. Paul,
{Phys. Rev. A} {\bf 52}, {(1995)} {R1801}.

\bibitem{Kiss1}
T. Kiss, U. Herzog, U. Leonhardt,
{Phys. Rev. A} {\bf 52}, {(1995)} {2433}. 
\end{thebibliography}
\end{document}